\documentclass[aps,prl,amsmath,amssymb,floatfix,twocolumn
]{revtex4-2}
\usepackage{hyperref}
\usepackage{graphicx}
\usepackage{xcolor}
\usepackage{bm}
\usepackage{braket}
\usepackage[ampersand]{easylist}
\usepackage[normalem]{ulem}
\usepackage{xfrac}
\usepackage{comment}

\newcommand{\cX}{\mathcal{X}}
\newcommand{\cZ}{\mathcal{Z}}

\begin{document}

\title{Obstruction to Ergodicity from Locality and $U(1)$ Higher Symmetries on the Lattice}

\author{Ramanjit Sohal}
\author{Ruben Verresen}
\affiliation{Pritzker School of Molecular Engineering, University of Chicago, Chicago, IL 60637, USA}

\date{\today}

\begin{abstract}
We argue that the presence of \emph{any} exact $U(1)$ higher-form symmetry, under mild assumptions, presents a fundamental obstruction to ergodicity under unitary dynamics in lattice systems with local interactions and finite on-site Hilbert space dimension. Focusing on the two-dimensional case, we show that such systems necessarily exhibit Hilbert space fragmentation and explicitly construct Krylov sectors whose number scales exponentially with system size. While these sectors cannot be distinguished by symmetry quantum numbers, we identify the emergent integrals of motion which characterize them. Our symmetry-based approach is insensitive to details of the Hamiltonian and the lattice, providing a systematic explanation for ergodicity-breaking in a range of systems, including quantum link models. 
\end{abstract}

\maketitle

\textit{Introduction.---}
Left to their own devices, ``typical" quantum systems are expected to thermalize \cite{DAlessio2016}. That is to say, 
when taken out of equilibrium, they equilibrate at late times under their own dynamics to states which appear thermal to correlation functions of local observables. 
Understanding the circumstances under which quantum many-body systems \emph{break} ergodicity is thus of fundamental conceptual interest, as this corresponds to a failure of the emergence of statistical mechanics. 
Advances in the development of quantum simulators \cite{QSim} have made the pursuit of this question a deeply practical one, as breaking ergodicity opens the possibility of protecting quantum information against the deleterious effects of thermalization.

One route to ergodicity breaking is provided by the phenomenon of Hilbert space fragmentation \cite{Sala2020,Khemani2020,Moudgalya2022}. %\rv{
This is a generalization of the familiar notion that if a Hamiltonian has a symmetry, then one can decompose the Hilbert space into Krylov sectors---subspaces which are dynamically disconnected, i.e., which do not evolve into one another under time evolution by said Hamiltonian. Indeed, sectors with distinct quantum numbers are in distinct Krylov sectors. Fragmentation is the remarkable scenario whereby the spatial locality of interactions ``fragments" a given quantum number sector into exponentially many Krylov sectors. This phenomenology has been identified in a range of systems \cite{Yang2020,Morningstar2020,Rakovzsky2020,Feldmeier2020,Langlett2021,Mukherjee2021,Khudorozhkov2022,Yoshinaga2022,Moudgalya2022comm,Lehmann2023,Li2023,Brighi2023,Kwan2025} and even been observed in experimental realizations of the Hubbard model  \cite{Scherg2021,Kohlert2023,Adler2024}.

Concomitantly, recent years have seen profitable study of 
\emph{higher-form} symmetries. 
Unlike standard, i.e. zero-form, global symmetries which act on all degrees of freedom of a system, higher-form symmetries act on spatial submanifolds \cite{Nussinov2009,Kapustin2017,Gaiotto2015,McGreevy2023}. Given that these symmetries lead to powerful constraints on low-energy physics, it is natural to expect they may have implications for non-equilibrium dynamics.
Indeed, recent works have demonstrated fragmentation in specific Hamiltonian and Floquet systems with particular higher-form symmetries \cite{Stephen2024,Stahl2024a,Khudorozhkov2024,Stahl2024b,Stahl2025,Orlov2025} and argued that the fragmentation remains robust for prethermal timescales under symmetry-breaking perturbations \cite{Stephen2024,Stahl2024a,Khudorozhkov2024,Stahl2024b,Stahl2025}.

In this work, we argue quite generally how the presence of higher-form symmetries leads to fragmentation on the lattice. In particular, we demonstrate a fundamental incompatibility between ergodicity and systems with local interactions, a finite on-site Hilbert space dimension, and exact $U(1)$ higher-form symmetries. For concreteness, we focus on 2D systems with one-form symmetries \footnote{By one-form symmetry, we mean specifically an exact, non-topological one-form symmetry. Here, exact means a symmetry which exactly commutes with the Hamiltonian, and does not arise only at low energy, and non-topological means that two one-form symmetry generators acting on two homologically equivalent loops need not have the same action on the Hilbert space.}.
We show that spatial loops of qudits with extremal $U(1)$ charge serve to separate distinct Krylov sectors, for which we construct explicit, emergent integrals of motion. Our results encompass many prior examples as special cases and provide a symmetry-based perspective on the dynamical constraints imposed by higher-form symmetries. 
We demonstrate our result in a series of increasingly general examples, before presenting our general argument.

\begin{figure}
    \centering
    \includegraphics[width=0.85\linewidth]{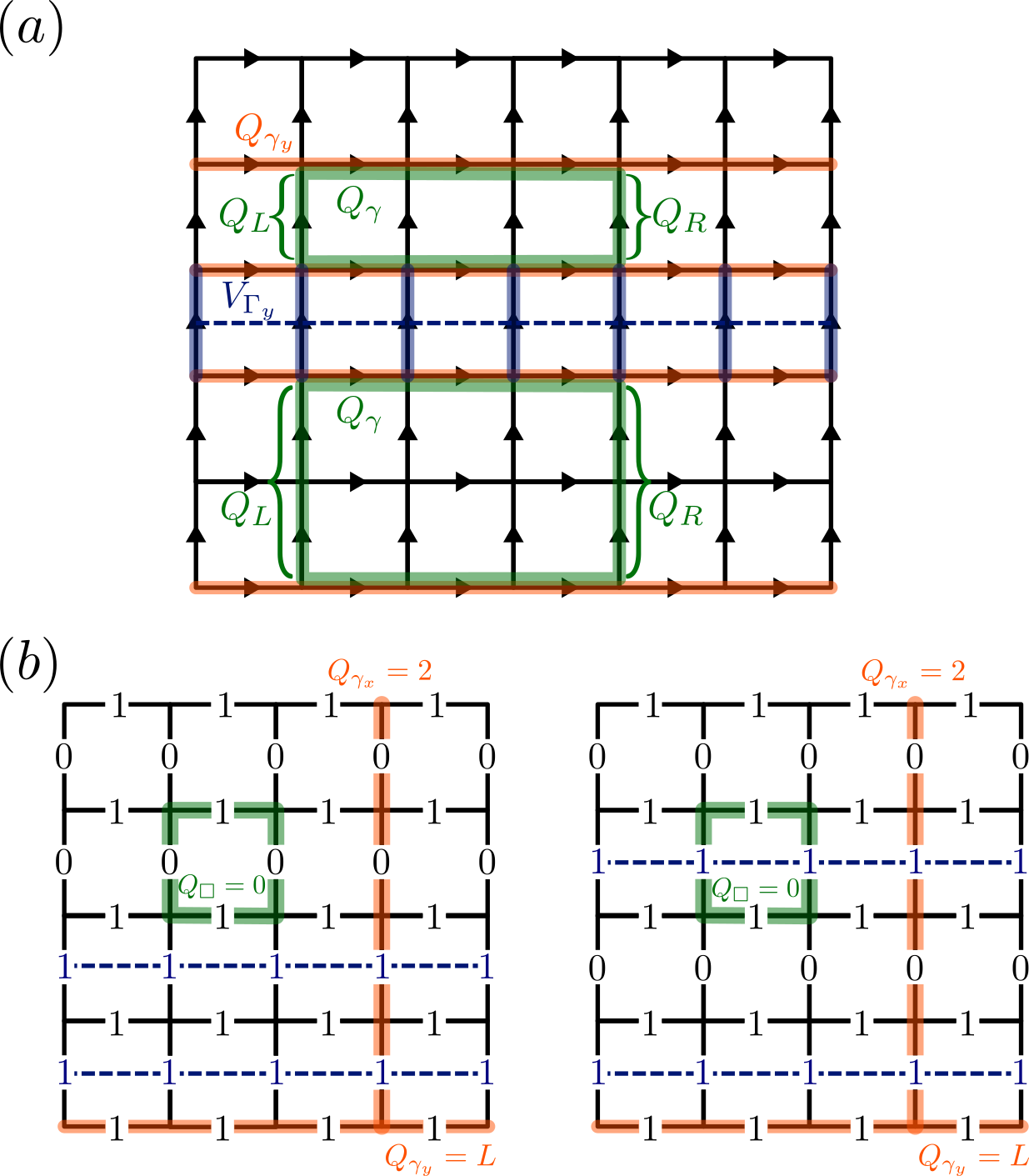}
    \caption{\textbf{Warm-up example.} (a) Oriented square lattice with top/bottom and left/right edges identified. Orange lines denote the one-form generators $Q_{\gamma_y}$ and the blue dashed line the charged operator $V_{\Gamma_y}$. Green loops denote $Q_\gamma$ on contractible loops and $Q_{L/R}$ the left/right edges of the corresponding $Q_\gamma$. (b) Two fragmented states with identical symmetry numbers under $Q_{\gamma_{x,y}}$ and all $Q_{\square}$, but lying in distinct Krylov sectors. The blue dashed lines indicate where $V_{\Gamma_y}$ were applied relative to the state with all spins up on the vertical bonds.
    }
    \label{fig:main}
\end{figure}

\textit{Warm-up with qubits.---} 
As a warm-up to illustrate our construction,  consider a system of qubits on the bonds of a square lattice (see Fig.~\ref{fig:main}) with periodic boundary conditions. We assign a direction to each edge and impose the $U(1)$ one-form symmetry,
\begin{align}
    U_\gamma(\theta) = \exp\left( i\theta\sum_{n \in \gamma} \alpha_n (1-Z_n)/2 \right) \equiv \exp(i\theta Q_\gamma) \, , \label{eq:qubit-symmetry}
\end{align}
where $\gamma$ is a closed loop on the direct lattice, $n$ labels consecutive sites along $\gamma$, and $\alpha_n=\pm 1$ if the orientation of $\gamma$ follows or goes against the direction of the edge $n$. Here, $X,Y,$ and $Z$ are the Pauli operators. We refer to $Q_\gamma$ as the $U(1)$ charge on the curve $\gamma$. For convenience, we set $q_n = \alpha_n (1-Z_n)/2 \in \{0, \pm 1\}$.
The two states depicted in Fig.~\ref{fig:main}(b) are distinct, yet have the same $U(1)$ one-form quantum numbers. We now introduce tools that allow us to generate exponentially many such sectors in a way that will generalize beyond this simple case, leading to fragmentation.
To do so, it will be useful to also have a ``raising operator" for this $U(1)$ one-form symmetry. To that end,
consider the charged operator $V_{\Gamma} = \prod_{n \in \Gamma} \sigma^{\beta_n}_n$, where $\Gamma$ is a closed, oriented curve on the \emph{dual} lattice; 
$\beta_n = \mathrm{sgn}(\hat{\Gamma}_n \times \hat{e}_n)$, where $\hat{\Gamma}_n$ is the direction of $\Gamma$ at site $n$, and $\hat{e}_n$ the direction of the lattice edge at $n$;
and $\sigma_n^{\pm}$ are the spin raising/lowering operators on site $n$. 
It is readily confirmed that $V_\Gamma$ has charge unity under the $U(1)$ one-form symmetry. In general, an operator $V_\Gamma$ has charge $k$ under a $U(1)$  one-form charge $Q_\gamma$, if $[Q_\gamma,V_\Gamma]$ is equal to $k$ times the signed intersection number between $\gamma$ and $\Gamma$; thus, $V_\Gamma$ increases the charge of $Q_\gamma$, if $\gamma$ and $\Gamma$ intersect once with the correct orientation, and $[V_\Gamma,Q_\square]=0$, where $\square$ is a plaquette.

Our claim is that any local Hamiltonian with the symmetry Eq.~\eqref{eq:qubit-symmetry} on this lattice necessarily exhibits Hilbert space fragmentation.
We demonstrate this by constructing the Krylov sectors, only making use of the symmetry itself and the charged operators $V_\Gamma$.
We first place each row $\gamma_y$ (with direction left to right) in an \emph{extremal} charge sector of $Q_{\gamma_y}$, i.e., $|Q_{\gamma_y}|$ is maximal [see top of Fig.~\ref{fig:main}(a)]. 
We will call any closed curve with maximal $|Q_{\gamma_y}|$ an \emph{extremal loop}.
This implies each $q_n$ takes an extremal value, fixing $Z_n = \mp 1$ if $Q_{\gamma_y}$ is maximized/minimized. Crucially, that the local generator \emph{can} be extremized is only possible due to the assumption of a finite on-site Hilbert space dimension. 

Now, in the subspace with all horizontal rows in fixed extremal states, let us consider a row, $\Gamma_y$, on the dual lattice and the charged operator $V_{\Gamma_y}$ which acts upon them. Let us place the spins on these bonds in a configuration such that $V_{\Gamma_y}$ does not annihilate them. In this case, there is a unique such configuration, in which all the bonds are placed in the spin-down state. Since the surrounding rows of horizontal bonds are fixed, this also fixes the values of the plaquette operators, $Q_{\square}$, along this row in the dual lattice. By construction, acting $V_{\Gamma_y}$ on this state yields another state with the \emph{same} $Q_{\square}$ quantum numbers [see Fig.~\ref{fig:main}(b)]. We can repeat this process on \emph{every} row of the dual lattice, yielding $2^{L_y}$ states, all with the same $Q_\square$ and $Q_{\gamma_y}$ quantum numbers. Now, consider the $U(1)$ one-form charge $Q_{\gamma_{x}}$ along some vertical row of the direct lattice, $\gamma_x$. Note that $Q_{\gamma_x'}$, for some other non-contractible vertical loop $\gamma_x'$, is related to $Q_{\gamma_x}$ via addition with some number of the $Q_\square$. 
While acting with $V_{\Gamma_y}$ will change the value of $Q_{\gamma_x}$, the latter can only take $2L_y+1$ values, and so cannot be used to distinguish all of the $2^{L_y}$ states.

Next, we construct a set of emergent integrals of motion labeling the Krylov sectors, implying they cannot be coupled by $U(1)$ one-form symmetric dynamics.
Let us consider a closed loop $\gamma$ encircling some number of plaquettes along a single row in one of the degenerate states [see Fig.~\ref{fig:main}(a)]. By construction, this symmetry generator has a value fixed by the enclosed plaquettes: $Q_{\gamma} = \sum_{\square \in \bar{\gamma}} Q_\square$, where $\bar{\gamma}$ indicates the region enclosed by $\gamma$. On the other hand, as the two surrounding horizontal rows are extremal, we can write $Q_{\gamma} = Q_L + Q_R + \mathrm{const.}$, where the constant is set by the values of $q_n$ on the horizontal bonds, while $Q_{L/R}$ are the generators on the left and right vertical bonds, respectively. 

We thus find that the sum $Q_L + Q_R$ is fixed to a constant value determined by the $Q_\square$ and the values of the $q_n$ on the horizontal lines, and hence is a constant of motion. However, by locality, we must in fact have that $Q_L$ and $Q_R$ are each \emph{separately} conserved, since we can take the length of the loop to be arbitrarily long. 
These operators are thus emergent integrals of motion within the subspace of extremal $Q_{\gamma_y}$. 
In fact, these integrals of motion are robust to arbitrary $k$-local deformations, as an interaction flipping the values of both $Q_L$ and $Q_R$ must also flip \emph{all} the vertical bonds along that row in order to preserve the $Q_\square$ quantum numbers; this is implemented by the $V_{\Gamma_y}$ operators. 

We have thus established that any Hamiltonian symmetric under Eq.~\eqref{eq:qubit-symmetry} exhibits  
an exponentially large in system size number of
Krylov sectors that cannot be distinguished by symmetry numbers, each labeled by an emergent integral of motion. In this case, each sector contains a single state with no entanglement, and thus also provide examples of quantum \emph{scars} \cite{Bernien2017,Shiraishi2017,Moudgalya2018,Turner2018,Chandran2023}.

\textit{Bulk fragmentation and general lattices.---}% 
We can construct Krylov sectors with more than a single state by considering non-dense packings of the lattice with extremal loops.
Indeed, suppose we only place some subset of the horizontal rows on the direct lattice in extremal eigenstates, leaving the states in the intervening horizontal rows undetermined.
The above argument runs through unchanged, where the emergent integrals of motion are now derived using a larger $Q_\gamma$ loop which connects the extremal lines, as in the bottom of Fig.~\ref{fig:main}(a).

\begin{figure}
    \centering
    \includegraphics[width=\linewidth]{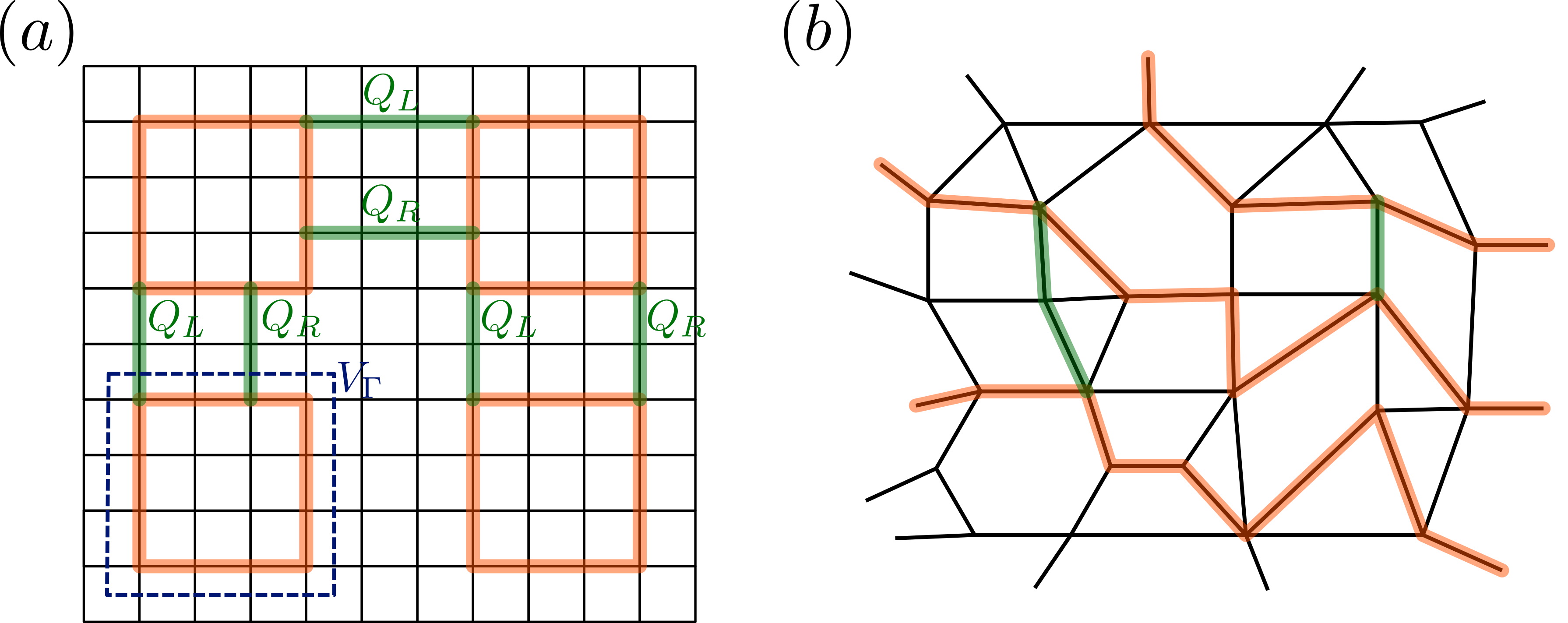}
    \caption{\textbf{Generalizing the warm-up example.} (a) Example of bulk fragmentation: we place the qubits along the orange lines in extremal states of $Q_{\gamma}$. The green lines indicate the emergent integrals of motion. 
    The dashed blue line is a charged operator toggling between the Krylov sectors. (b) Fragmentation on a general lattice. Orange lines indicate strings of qubits in extremal eigenstates of $Q_\gamma$ and green lines indicate the emergent integrals of motion.
    }
    \label{fig:bulk}
\end{figure}

Building on this, we can show $U(1)$ one-form symmetry in fact implies \emph{bulk} fragmentation---Krylov sectors associated to \emph{contractible} extremal loops.
Indeed, consider a configuration of 
contractible extremal loops, as in Fig.~\ref{fig:bulk}(a). 
Following the same arguments as above, we conclude that we have emergent integrals of motions associated to open $U(1)$ one-form symmetry generators connecting pairs of these contractible loops. 
These Krylov sectors are stable to $k$-local perturbations for $k$ less than the length of the extremal loops; indeed, it is the action of a $k$-local operator, $V_\Gamma$ for a loop $\Gamma$ encircling an extremal loop, that toggles between Krylov sectors.

Finally, it is clear that our Krylov sector construction in fact holds on general lattices for which a one-form symmetry structure can be consistently defined. 
We do not require the lattice to be regular [see Fig.~\ref{fig:bulk}(b)].
To generate a set of Krylov sectors, one simply places some set of loops on the direct lattice in extremal states. Following the above discussion, emergent integrals of motion labeling the Krylov sectors are given by the truncated one-form symmetry operators connecting these extremal loops. 
Moreover, while we have focused on 2D systems for concreteness, our approach is readily extended to higher dimensions, where the symmetry and/or charged operators will now act on higher-dimensional surfaces.

\textit{General on-site $U(1)$ generators.---} 
We have established that any lattice system symmetric under Eq.~\eqref{eq:qubit-symmetry} exhibits fragmentation. The same argument implies any lattice system with general on-site Hilbert space dimension and a $U(1)$ one-form symmetry with a strictly on-site generator also necessarily exhibits fragmentation.
Let us consider a lattice system of $d$-dimensional qudits on each bond. 
On each edge, we define the generalized Pauli operators $\cX_e$ and $\cZ_e$, obeying the algebra $\cZ_e \cX_e = \omega \cX_e \cZ_e$ with $\omega = \exp(2\pi i / d)$. Let us define the projector onto the $k^{th}$ state of the qudit on edge $e$ as $P_e^k = \ket{k}_e\bra{k}_e$, where $\cZ_e\ket{k}_e = \omega^k \ket{k}_e$.
We consider the $1$-form symmetry generated by,
\begin{align}
    U_\gamma(\theta) = \exp\left(i\theta \sum_{n\in \gamma} q_n \right), \quad q_n = \alpha_n \sum_{k=0}^{d-1} \lambda_k P_n^k \, , \label{eq:qudit-symmetry}
\end{align}
where $\lambda_k \in \mathbb{Z}$ to ensure the generator has integer spectrum and $\alpha_n$ is defined below Eq.~\eqref{eq:qubit-symmetry}. This constitutes the most general family of $U(1)$ one-form symmetries with strictly on-site generator. 
One can repeat the same steps as in the qubit example to demonstrate that fragmentation necessarily follows from the imposition of this symmetry; we provide the details in the End Matter.

\begin{figure}
    \centering
\includegraphics[width=0.85\linewidth]{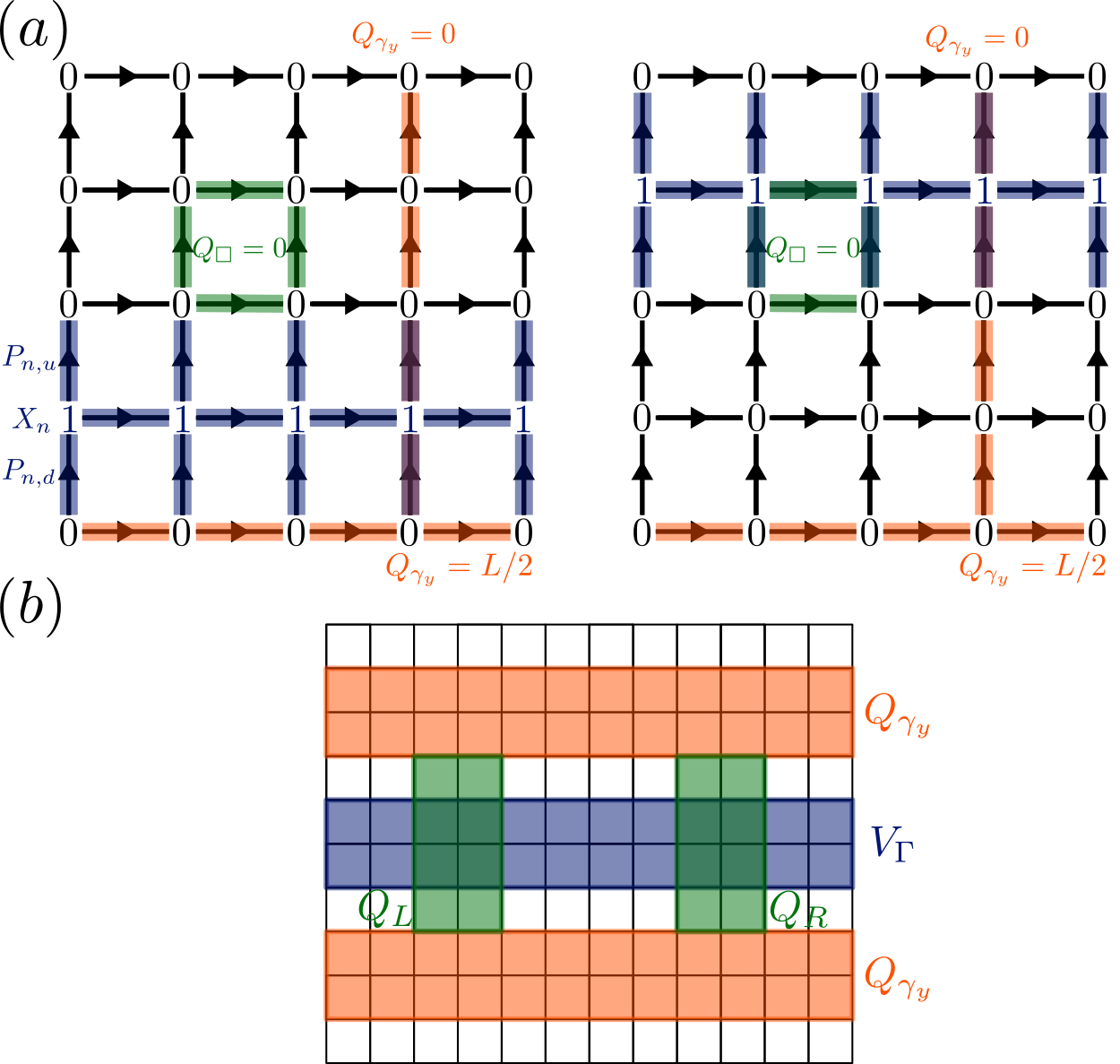}
    \caption{\textbf{Fragmentation for non-on-site $U(1)$ generators.} (a) Examples of states in distinct Krylov sectors in a square lattice system with the symmetry Eq.~\eqref{eq:ising-symmetry}. The shaded blue ribbons indicate where the charged operators $V_\gamma$ were applied. (b) Fragmentation from general one-form symmetry on the square lattice. The orange, green, and blue strips indicate the symmetry generators, emergent integrals of motion, and charged operators, respectively.
    }
    \label{fig:ising}
\end{figure}

\textit{Non-on-site example.---} 
Before proceeding to the most general case, we first discuss an example in which the $U(1)$ one-form symmetry generator is \emph{not} on-site.
Let us consider an oriented square lattice with periodic boundary conditions, but with qubits placed on the vertices [see Fig.~\ref{fig:ising}(a)], and impose the $U(1)$ one-form symmetry,
\begin{align}
    U_\gamma(\theta) = \exp\left(i\theta \sum_{n\in \gamma} q_n \right), \quad q_n = \alpha_n \frac{1+Z_n Z_{n+1}}{4} \, . \label{eq:ising-symmetry}
\end{align}
Despite the generator being non-on-site, we claim that any Hamiltonian with this symmetry must exhibit fragmentation. 

The argument proceeds in essentially the same fashion as the preceding examples. We place each horizontal row, $\gamma_y$, on the direct lattice in an extremal eigenstate of $Q_{\gamma_y}$---for concreteness, let us choose the maximal eigenstates. This amounts to setting $q_n = Z_n Z_{n+1} = 1/2$ for each $n$ in a given row $\gamma_y$. There is a two-dimensional subspace of states satisfying this constraint, spanned by the all-spin up and down states, $\ket{00\dots 0}$ and $\ket{11 \dots 1}$, respectively. 
As before, in this subspace, we have for a closed $\gamma$ encircling some number of consecutive plaquettes along a row that $Q_\gamma = Q_L + Q_R + \mathrm{const.}$, with the constant set by the $q_n$ on the surrounding rows of qubits on the direct lattice. From the arguments in the preceding examples, we again conclude that the $Q_{L/R}$ are emergent integrals of motion. 
We note that in this case, $Q_{L/R}$ label whether two consecutive layers are (anti)aligned.%}

In order to show that the corresponding Krylov sectors are non-empty and the states they contain are not distinguished by local symmetry quantum numbers, we require a charged one-form operator. One may verify that the following operators have charge unity under the one-form symmetry: $V_{\gamma_y} = \prod_{n \in \gamma_y} X_n P_{n,u}^+ P_{n,d}^+$. 
Here, $\gamma_y$ is a closed horizontal path on the \emph{direct} lattice, $u/d$ label the qubits immediately above and below $n$, and $P_{i,j}^+ = (1 + Z_i Z_j)/2$ is a projector onto the ferromagnetic configurations of qubits $i$ and $j$ [see Fig.~\ref{fig:ising}(a)]. 

\textit{General $U(1)$ one-form symmetries.---}
We now discuss the most general case and present a sufficient set of assumptions on the $U(1)$ generator which will guarantee fragmentation.
Let us consider an arbitrary lattice system with finite, on-site Hilbert space dimension and $U(1)$ one-form symmetry generated by,
\begin{align}
    U_\gamma(\theta) = \exp\left(i \sum_{n\in \gamma} \alpha_n q_n \right) \, . \label{eq:general-symmetry}
\end{align}
We do not require the $q_n$ to be translation invariant, beyond requiring that $U_\gamma(\theta) = U_{\gamma_1}(\theta)U_{\gamma_2}(\theta)$ for closed loops $\gamma$, $\gamma_1$, and $\gamma_2$ such that $\gamma = \gamma_1 + \gamma_2$. We further assume the existence of a set of charge-$k$ one-form operators, $V_\Gamma(k)$, with some charges $k \in \mathbb{Z}$ under the $U(1)$. 

We make one assumption regarding the one-form symmetry generator. Consider a closed curve $\gamma$. Crucially, assuming a finite on-site Hilbert space dimension means 
that $Q_\gamma$ is bounded above and below. 
Now take a state $\ket{\psi}$ for which $Q_\gamma$ obtains an extremal value. We \emph{assume} this implies that $\ket{\psi}$ is also an eigenstate of the $q_n$ (for some choice of local generator $q_n$), i.e., $q_n \ket{\psi} = \kappa_n \ket{\psi}$ for some $\kappa_n \in \mathbb{R}$. 
This is automatically satisfied if the $q_n$ are commuting projectors, which is in fact natural for a $U(1)$ generator.
Physically, this assumption expresses the expectation that maximizing the total $U(1)$ one-form charge should come from maximizing the local charge densities. 
Indeed, we are not aware of any examples of $U(1)$ one-form charges that would not satisfy this assumption.

We can repeat our earlier arguments to construct Krylov sectors on the torus. We again use extremal states on horizontal lines $\gamma_y$, the only difference being that the generators can now have a width, such that we space out these lines [orange regions in Fig.~\ref{fig:ising}(b)]. The construction of local integrals of motion is also similar to before, leading to fattened $Q_{L/R}$ operators [green regions in Fig.~\ref{fig:ising}(b)], which are toggled by acting with the charged string operator.
The argument for bulk fragmentation likewise follows identically, with the requirement that the extremal loops $\gamma$ should be chosen such that the $Q_\gamma$ have non-overlapping support. This establishes the claimed result of fragmentation in any exact $U(1)$ one-form symmetric lattice Hamiltonian with local interactions and finite on-site Hilbert space dimension.

\begin{figure}
    \centering
    \includegraphics[width=0.9\linewidth]{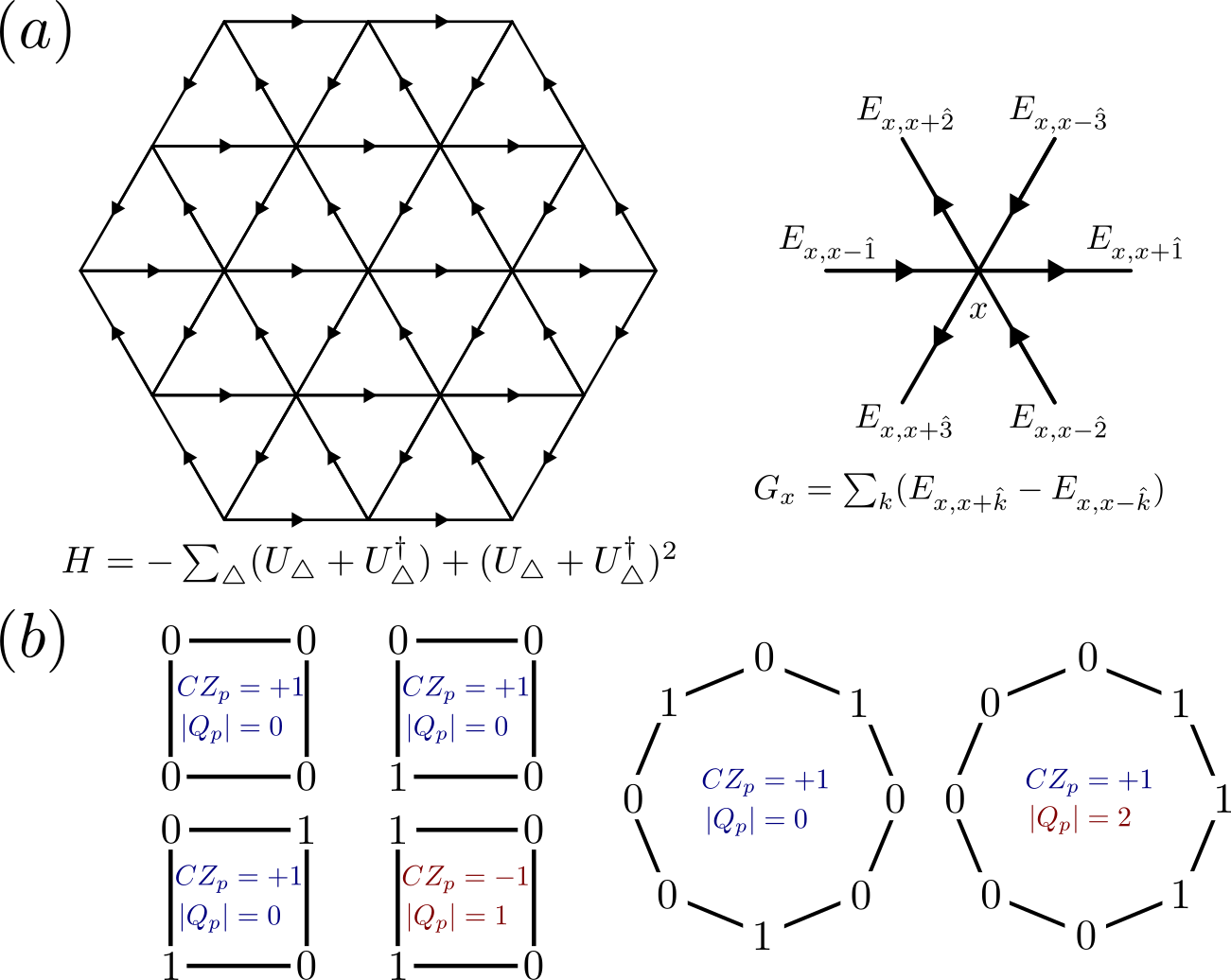}
    \caption{\textbf{Applications.} (a) Quantum link model on the triangular lattice with $U_\triangle = \prod_{l \in \triangle} U_l$ (left) and Gauss' law operator (right). (b) Examples of qubit configurations on a square plaquette indicating the presence of a $U(1)$ one-form charge only in the presence of a CZ one-form charge and configurations on octagonal plaquettes with trivial CZ charge but non-trivial $U(1)$ one-form charge.
    }
    \label{fig:applications}
\end{figure}

\textit{Applications.---} 
A natural setting for the application of our results is that of \emph{quantum link models} (QLMs). These realize embeddings of $U(1)$ gauge theories in lattice spin systems \cite{Chandrasekharan97,Wiese21} and thus
provide natural models for the study of gauge theories in quantum simulators \cite{halimeh2025}.
Fig.~\ref{fig:applications}(a) depicts a QLM on an oriented triangular lattice, with qubits on the links. The electric fields and gauge connections are realized in terms of the spin-1/2 operators on the link $l$ as $E_l = S_l^z$ and $U_l = S_l^+$, such that 
$[E_l,U_{l'}]=i\delta_{l,l'}U_l$. 
The Hamiltonian is required to commute with the Gauss' law [see Fig.~\ref{fig:applications}(a)], 
which is the generator of a $U(1)$ one-form symmetry on the dual lattice. 
Our results thus immediately imply the existence of scar states and fragmentation in any such QLM Hamiltonian. 
Indeed, there is now a substantial literature investigating 
violations of thermalization
in lattice gauge theories generally \cite{Karpov2021,Halimeh2023,Budde2024,Jeyaretnam2025,Ciavarella2025,halimeh2025}, including scar states in 2D QLMs specifically \cite{Banerjee2021,Biswas2022,Sau2024}, 
based primarily on empirical numerical evidence and studies of specific Hamiltonians. 
Our work provides, to our knowledge, the first systematic explanation for ergodicity breaking via fragmentation in 2D QLMs. %\raman{Mention connections to quantum dimer models}
Our symmetry-based arguments also apply to higher dimensions, agreeing with the fragmentation observed in a $3d$ QLM model studied in Ref.~\onlinecite{Steinegger2025}.

While an exact $U(1)$ one-form symmetry seems like a strong constraint to impose, such a symmetry can arise by only imposing a \emph{discrete} one-form symmetry in certain lattices. For instance, consider a lattice with qubits on vertices and the ``CZ" (controlled-Z) one-form symmetry, $\mathrm{CZ}_\gamma = \prod_{n\in \gamma} \mathrm{CZ}_{n,n+1}$, where $\mathrm{CZ}\ket{a,b} = (-1)^{ab}\ket{a,b}$ \cite{Stephen2024,Zhang2026}. This generates the $\mathbb{Z}_2$ subgroup of a $U(1)$ symmetry generated by $Q_\gamma=\sum_n(-1)^nZ_n Z_{n+1}/4$, i.e., $U_\gamma(\pi) = \mathrm{CZ}_\gamma$. A Hamiltonian symmetric under the CZ symmetry will \emph{necessarily} be symmetric under this larger $U(1)$ symmetry if, for any given plaquette $p$, the $\mathbb{Z}_2$ charge vanishes, $\mathrm{CZ}_p = +1$, if and only if the $U(1)$ charge vanishes, $Q_p = 0$. By enumerating all spin configurations, one sees this is the case for square and hexagonal plaquettes, but not octagonal plaquettes [see Fig.~\ref{fig:applications}(b)]. Thus, a square or hexagonal lattice Hamiltonian with the $\mathrm{CZ}$ one-form symmetry automatically is symmetric under the larger $U(1)$ one-form symmetry (akin to the pivot mechanism in Ref.~\onlinecite{Tantivasadakarn23}) 
and thus will exhibit fragmentation. 

This provides a symmetry-based perspective on the fragmentation of the $\mathrm{CZ}_p$ model of Ref.~\onlinecite{Stephen2024}, a Hamiltonian on the square lattice with the $\mathrm{CZ}$ one-form symmetry. 
Our approach allows for identifying the same scar states constructed via the microscopic approach in Ref.~\onlinecite{Stephen2024}, as well as the coarser and bulk Krylov sectors discussed above. 
Moreover, our approach suggests the fragmentation of the ${\rm CZ}_p$ model should not apply to other lattices such as the square-octagon lattice, as they lack a $U(1)$ one-form symmetry [Fig.~\ref{fig:applications}(b)]. 
The enhancement of a discrete one-form to a $U(1)$ one-form symmetry can occur more generally for non-on-site one-form symmetries; such systems thus present natural platforms for the study of fragmentation.

Separately, Ref.~\onlinecite{Stahl2024a} studied square lattice systems with multiple $U(1)$ one-form symmetries, of which Eq.~\eqref{eq:qudit-symmetry} forms a subgroup. The authors of Ref.~\onlinecite{Stahl2024a} argued for fragmentation by noting the flux-free Hilbert space may be understood as describing states of non-intersecting loop configurations. In contrast, our symmetry-based arguments do not make explicit reference to the microscopic structure of the frozen sectors, also yield bulk fragmentation, and are not restricted to the square lattice.

Our symmetry-based analysis further allows for arguing the presence of fragmentation in a larger class of systems through \emph{gauging}. For instance, gauging the $\mathbb{Z}_2$ subgroup of the $U(1)$ one-form symmetry of Eq.~\eqref{eq:qubit-symmetry}, generated by $U_\gamma(\pi)=\prod_i X_i$, yields a system with the symmetry Eq.~\eqref{eq:ising-symmetry}. 
States in the sector with trivial $\mathbb{Z}_2$ one-form plaquette charges survive the gauging (other states map to defect sectors of the gauged theory) and remain indistinguishable with respect to the new one-form charges. Since this sector exhibited fragmentation, this also establishes fragmentation in the gauged system.

More generally, given a system with a $U(1)$ one-form symmetry to which our arguments apply, gauging a subgroup of this $U(1)$ or any other finite global symmetries present will yield another system potentially exhibiting fragmentation.
In particular, consider a system symmetric under Eq.~\eqref{eq:qubit-symmetry} and the global spin-flip symmetry $X = \prod_i X_i$, which sends $Q_\gamma \mapsto - Q_\gamma$. Gauging the spin-flip symmetry yields a system with a \emph{non-invertible} $U(1)$ one-form symmetry, with ``cosine" symmetry  $U_\gamma(\theta) + U_\gamma(-\theta)$ \cite{Bhardwaj23,Hsin2024}, which we thus expect to exhibit fragmentation. 

\textit{Conclusions.---}We have argued for a fundamental obstruction to ergodic dynamics in lattice systems with local interactions and exact $U(1)$ higher-form symmetries. Our arguments are purely symmetry based, lattice-independent, and show that the algebra of the $U(1)$ one-form symmetry generators and charged operators, making a minimal set of assumptions on the former, necessarily leads to Hilbert space fragmentation.
While we focused on 2D systems for concreteness, it is clear that our arguments can be readily extended to higher dimensions, though it would be important to explore this in concrete models. In particular, we leave open the question of whether our mechanism can be generalized to when the $U(1)$ one-form symmetry is not \textit{faithful}---i.e., when there do not exist charged states for the non-contractible $U(1)$ one-form generator \cite{Hsin2026}. Likewise, it would be interesting to see if similar arguments can be applied to systems with \textit{subsystem symmetries}; our construction of local integrals of motion relied crucially on the symmetry generator being deformable.
It would also be important to study in future work  the effects of breaking the higher symmetry---see the End Matter for comments along this direction.
It would also be interesting to explore the implications of the spectral constraints obtained here on the emergence of hydrodynamics and on entanglement dynamics. 

\textit{Acknowledgments.---} We are grateful to Pranay Gorantla, Emilie Huffman, Tsung-Cheng Lu, Pablo Sala, and David Stephen for helpful discussions. We thank KITP for hospitality during the program ``Noise-robust Phases of Quantum Matter", during which part of this work was completed. This research was supported in part by grant NSF PHY-2309135 to the Kavli Institute for Theoretical Physics (KITP).

\bibliography{frag-references}

\appendix

\section{End Matter}

\subsection{Details of General on-site Hilbert Space Dimension Example}
Here, we provide some additional details on the argument for fragmentation enforced by the symmetry of Eq.~\eqref{eq:qudit-symmetry} in qudit systems. We focus on the maximally fragmented sectors on the square lattice with periodic boundary conditions; the arguments for bulk fragmentation and fragmentation on general lattices are simple extensions of this.

As before, we place each horizontal row on the direct lattice in an extremal eigenstate of $Q_\gamma$.  
As the charge is on-site, this amounts to extremizing $q_n$ on each bond; note that there is not a unique such state if the extremal $\lambda_k$ are degenerate. This does not affect our argument---the crucial feature is that extremizing $Q_\gamma$ uniquely fixes the value of $q_n$ on each bond. 
If we consider a closed loop $\gamma$ on the direct lattice formed by a contiguous set of plaquettes between the consecutive horizontal rows, we see that $Q_\gamma = Q_L + Q_R + \mathrm{const.}$, where the constant is set by the values of $q_n$ on the extremal rows. The $Q_{L/R}$ are again robust integrals of motion, 
labeling distinct Krylov sectors. 
Any degeneracies of the $\lambda_k$ will lead to degeneracies in the spectrum of $Q_{L/R}$, which correspond to the dimensions of the Krylov sectors.
Note that if, say, $q_L = \lambda_i$ where $\lambda_i = \lambda_j$ for some $i \neq j$, then we immediately have that the corresponding Krylov sector is at least two-dimensional. Indeed, any vertical bond that takes on a degenerate eigenvalue contributes a corresponding factor of the degeneracy to the size of the Krylov sector.

We can also construct distinct states in a given Krylov sector through the use of the charge-$k$ one-form operators
%\begin{align}
    $V_\Gamma(k) = \prod_{n\in \Gamma} \Sigma_n^{\beta_n}(k)$,
%\end{align}
where $\Sigma_n^+(k) =\ket{k}_n\bra{k_0}_n$, with $P_n^l \ket{k} = \delta_{k,l}\ket{k}$, is the raising operator on site $n$ and $\Sigma_n^-(k) \equiv (\Sigma_n^+(k))^\dagger$. 
With these operators in hand, we can construct states distinguished only by the emergent integrals of motion by placing a given row, $\Gamma_y$, on the dual lattice in a state such that it is not annihilated by (at least one of) the $V_{\Gamma_y}(k)$; in the present case, this corresponds to placing each qudit in the state $\ket{k_0}$. The fragmented states are then obtained by acting with the $V_{\Gamma_y}(k)$ for each $k$. This establishes the presence of the maximally fragmented Krylov sectors on the square lattice. 

\subsection{Comments on Stability}

It is natural to ask whether the fragmentation that follows from our arguments in a $U(1)$ one-form symmetric system persists under a perturbation breaking this symmetry. This question was answered in the affirmative for specific examples of Hamiltonians exhibiting exact higher-form symmetries in Refs.~\onlinecite{Stephen2024,Stahl2024a,Khudorozhkov2022,Stahl2024b}. Rigorously establishing stability in the general setting of the systems of our interest is beyond the scope of this work, but we can leverage some of the techniques employed in Refs.~\onlinecite{Stephen2024,Stahl2024a,Khudorozhkov2022,Stahl2024b} to give a suggestive argument in favor of stability.

Consider a $U(1)$ one-form symmetric lattice Hamiltonian $H_0$ to which our arguments apply, such that it exhibits fragmentation. Let the $U(1)$ one-form symmetry be generated by $Q_\gamma$. We are interested in understanding whether $H = H_0 + V$, where $V$ is some interaction which explicitly breaks the $U(1)$ one-form symmetry, still exhibits some signatures of fragmentation. To that end, let us consider a \emph{different} Hamiltonian, $H' = H + \lambda \sum_p Q_p$, where the summation is over all plaquettes $p$ and $Q_p$ is the $U(1)$ generator around the plaquette $p$. The Hamiltonian $Q \equiv \sum_p Q_p$ is trivially symmetric under the same $U(1)$ one-form symmetry as $H_0$ and thus exhibits the same Krylov sectors. The key point is that, as a sum of local $U(1)$ generators, $Q$ has an integer valued spectrum. In the limit of large $\lambda$, we can treat $H'$ as $\lambda Q$ perturbed by $H$. As in Ref.~\onlinecite{Stephen2024}, we can then invoke the theory of prethermalization \cite{Abanin2017}, which states that perturbation theory of a Hamiltonian with integer spectrum only breaks down (i.e. the Krylov sectors become coupled) at time scales exponentially large in $\lambda$. 

This argument implies that the fragmentation of a $U(1)$ one-form symmetric Hamiltonian $H_0$ can be \emph{made} stable (in the sense that it will persist to prethermal time scales) by adding the local generators of the $U(1)$ one-form to the Hamiltonian. As a corollary, this implies that Hamiltonians constructed from local $U(1)$ one-form generators will exhibit, in this sense, stable fragmentation under symmetry-breaking perturbations. We leave to future work a more systematic understanding of the stability of the fragmentation in $U(1)$ one-form symmetric systems in the general setting.

\end{document}